\documentclass[12pt,a4paper]{article}   
\usepackage[dvips]{graphicx}

\def\m@th{\mathsurround=0pt}

\def\fsquare(#1,#2){
\hbox{\vrule$\hskip-0.4pt\vcenter to #1{\normalbaselines\m@th
\hrule\vfil\hbox to #1{\hfill$\scriptstyle #2$\hfill}\vfil\hrule}$\hskip-0.4pt
\vrule}}

\newcommand{\np}{\nonumber\\}
\newtheorem{thm}{Theorem}
\newtheorem{lem}{Lemma}
\newtheorem{prop}{Proposition}

\newtheorem{df}{Definition}
\newtheorem{cj}{Conjecture}
\def\dl{\frac{1}{10}}
\def\sq{\sqrt{2}}
\def\asy{\pm\infty}

\def\renv{\frac{v}{10}}

\title{
Quantum Jacobi-Trudi Formula
and $E_8$ Structure in the Ising Model in a Field
}
\author{ J. Suzuki\thanks{e-mail: suz@hep1.c.u-tokyo.ac.jp}\\
        \parbox{0.9\textwidth}{
        {\em
        \begin{center}
       Institute of Physics,\\
        University of Tokyo at Komaba,\\
       Komaba 3-8-1, Meguro-ku, Tokyo,\\
       Japan
        \end{center}
        }}
       }
\date{March 1998}
\begin{document}
\maketitle
\begin{abstract}
  We investigate a 1D quantum system associated with 
  the Ising model in a field(the
  dilute $A_3$ model) by the recently developed quantum 
  transfer matrix (QTM) approach.
 A  closed set of functional relations is found among variants of
  fusion  QTMs which are characterized by skew Young tableaux.
  These relations are proved by using
  a quantum analogue of Jacobi-Trudi formula, 
  together with special features at
  "root of unity" .
  The numerical analysis on their eigenvalues shows 
  a remarkable coincidence with exponents
  characteristic to $E_8$.
  From these findings, we have successfully recovered 
  the $E_8$ Thermodynamic Bethe ansatz equation 
  by Bazhanov et al, however, without specific choice 
  of strings solutions. 
\end{abstract}
\clearpage

\section{Introduction}

Perhaps the most striking prediction from the
deformation of conformal field theory to
statistical mechanics is the hidden $E_8$ 
universality structure in the Ising model 
in a field \cite{Zam891,Zam892}.
This triggers active researches in both 
field theory\cite{HenkSal}-\cite{DelfMuss}
and statistical mechanics \cite{WNS1}-\cite{MO97}.
In this report, we shall attack the problem
in view of a solvable lattice model .

In \cite{WNS1}, a  $L-$ state RSOS model is proposed of which 
Boltzmann weights are parameterized by elliptic functions which
satisfy the Yang-Baxter relations \cite{Yang67}-\cite{Baxbook}.
The model, hereafter referred to as the dilute $A_L$ model,
 has  the outstanding property.
For odd $L$'s, the up-down symmetry of the local state is
broken for nonzero elliptic nome .
One may thus identify the magnetic field  with  elliptic nome.
Actually, this identification leads to
the desired critical exponent $\delta=15$ 
of the Ising model for
$L=3$ \cite{WNS1, WNS2, WPNS}.
There are further supports for the conjecture that 
the dilute $A_3$ model
belongs to the same universality class with the Ising model in 
a field, say, the central charge, critical exponents for bulk and
a system with boundaries and so on \cite{WNS1}-\cite{BS2}.

From these two observations, one may 
naturally expect the $E_8$ structure
behind the  dilute $A_3$ model.
As an evidence, we refer to the Thermodynamic Bethe Ansatz(TBA)
approach in \cite{BWN}.
These authors deal with the one dimensional counterpart to
the  dilute $A_L$ model, and discuss the finite
temperature problem.
They claim the dominating nine patterns (strings) of Bethe
ansatz roots for $L=3$ in the thermodynamics limit, and show that the
string hypothesis  leads to the $E_8$ structure for the TBA equation.
This observation originates from the numerical investigation 
near the zero elliptic nome.
These strings are, however, quite involved and could 
lead to numerical difficulty with 
increase in the elliptic nome \cite{GN1,GN2}.
It may be preferable to have
the derivation of $E_8$ structure free from
specific choice of strings.

In this report, we analyze the same problem
through a  different route based on the quantum
transfer matrix (QTM) \cite{Suz85,Klu92}.
The QTM approach has been developed recently
as an alternative method in studies of 1D quantum systems
at finite temperatures \cite{Suz85}-\cite{KSS98}.
The formulation yields the expression of the free energy by the
largest eigenvalue of the object called QTM.
The method has some advantages over the standard approach 
\cite{Gaudin71}-\cite{TakSuz}
based on the string hypothesis.
For example, it yields a transparent derivation of 
the correlation lengths at finite temperatures \cite{KSS98, BLZ96, Fend97}.
Once combined with the Yang-Baxter integrability structure,
the evaluation of physical quantities at $T>0$ reduces
to the study on analyticity of some auxiliary functions
\cite{Klu92,Klu93},\cite{JK96}-\cite{KSS98}.
The crucial point in this approach lies in the choice of the
auxiliary function, which is not necessary unique.
One may adopt convenient functions having nice analytic properties.
Here  we choose them utilizing a family of commuting
transfer matrices which includes QTM as the most fundamental one.
We call them fusion QTMs.
The functional relations among them allow the explicit
evaluation of their eigenvalues.
This strategy has been successfully applied to several models
\cite{Klu92, JKSfusion, KSS98}.
There fusion QTMs come into play which corresponds to Yangian modules 
 specified by {\it only}  Young tableaux of rectangle shape.
For the present model, this is no longer true.
They constitute  closed functional relations, however,
lack nice analytic property.
The absence of the property prevents the derivation of
integral equations (TBA).
From trials and errors, we find it necessary to 
introduce  variants of QTMs corresponding to
"skew shape" Young tableaux. 
A quantum analogue of Jacobi-Trudi formula 
\cite{BR, KS, KOS, AK} as well as
special features at "root of unity" 
 play fundamental roles in the derivation of the
 functional relations.
Numerical investigations on analyticity of "modified" QTMs show a 
remarkable coincidence of imaginary parts of locations of zeros
with the exponents characterizing massive particles
in $E_8$ scattering theory \cite{Dorey}.
This validates a modest assumption on
the existence of a strip in the complex plane 
where "modified" QTMs  are free from zeros and singularities.
Based on these, we successfully recover the TBA equation
from the functional relations, namely, 
totally independent of string hypothesis.

This paper is organized as follows.
In the next section, we give a brief review on the dilute
$A_L$ models.
A minimal guide to the QTM approach is sketched in section 3.
The $sl_3$ fusion structure found in \cite{ZPG}
will be explained in section 4.
Based on these preparations, we introduce fusion QTMs parameterized
by skew Young tableaux  and their variants in section 5.
The latter are found to 
satisfy closed functional relations.
Section 6 is devoted to the observation of the 
analyticity of these functions.
Assuming Conjecture 1 and 2, 
the $E_8$ TBA equation is derived from the functional relations.
We conclude the paper with brief summary and discussion in section 7.
%
%
\section{dilute $A_L $ model}
The dilute $A_L$ model is proposed in \cite{WNS1} as
an elliptic extension of the Izergin-Korepin model \cite{IK}.
The model is of the restricted SOS type with local 
variables $\in \{1,2,\cdots,L \}$.
The variables $\{a, b\} $ on neighboring sites
should satisfy adjacency condition, $|a-b|\le 1$.
The solvable weights are given by,
\begin{eqnarray}
\raise 2mm \vtop{\hbox{$a$}\hbox{$a$}}
\,\framebox[0.4cm][c]{$u$} \,
\raise 2mm \vtop{\hbox{$a$}\hbox{$a$}}
&=& \frac{\theta_1(6-u)\theta_1(3+u)}{\theta_1(6)\theta_1(3)}- 
    \frac{\theta_1(u)\theta_1(3-u)}{\theta_1(6)\theta_1(3)} \times \np
& & \Bigl(
           \frac{S_{a+1}}{S_a} \frac{\theta_4(2a-5)}{\theta_4(2a+1)}
          +\frac{S_{a-1}}{S_a} \frac{\theta_4(2a+5)}{\theta_4(2a-1)}
    \Bigr ), \np
\raise 2mm \vtop{\hbox{$a\pm 1$}\hbox{$\quad a$}}
\,\framebox[0.4cm][c]{$u$} \>
\raise 2mm \vtop{\hbox{$a$}\hbox{$a$}}&=&
\raise 2mm \vtop{\hbox{$a$}\hbox{$a$}}
\> \framebox[0.4cm][c]{$u$} \>
\raise 2mm \vtop{\hbox{$a$}\hbox{$a\pm 1$}}
=\frac{\theta_1(3-u)\theta_4(\pm 2a+1-u)}{\theta_1(3)\theta_4(\pm 2a +1)},
\np
\raise 2mm \vtop{\hbox{$\quad a$}\hbox{$a\pm 1$}}
\,\framebox[0.4cm][c]{$u$} \>
\raise 2mm \vtop{\hbox{$a$}\hbox{$a$}}&=&
\raise 2mm \vtop{\hbox{$a$}\hbox{$a$}}
\> \framebox[0.4cm][c]{$u$} \>
\raise 2mm \vtop{\hbox{$a\pm 1$}\hbox{$a$}}
=\Bigl ( \frac{S_{a\pm 1}}{S_a} \Bigr )^{1/2}
\frac{\theta_1(u)\theta_4(\pm 2a-2+u)}{\theta_1(3)\theta_4(\pm 2a +1)},
\np
\raise 2mm \vtop{\hbox{$a$}\hbox{$a$}}
\> \framebox[0.4cm][c]{$u$} \>
\raise 2mm \vtop{\hbox{$a\pm 1$}\hbox{$a\pm 1$}}&=&
\raise 2mm \vtop{\hbox{$a\pm 1$}\hbox{$\quad a$}}
\> \framebox[0.4cm][c]{$u$} \>
\raise 2mm \vtop{\hbox{$a\pm 1$}\hbox{$\quad a$}}
=\Bigl ( \frac{\theta_4(\pm 2a +3)\theta_4(\pm 2a-1)}
              {\theta_4^2(\pm 2a+1)} \Bigr )^{1/2}
\frac{\theta_1(u)\theta_1(3-u)}{\theta_1(2)\theta_1(3)},
\np
\raise 2mm \vtop{\hbox{$a\pm 1$}\hbox{$\quad a$}}
\> \framebox[0.4cm][c]{$u$} \>
\raise 2mm \vtop{\hbox{$\quad a$}\hbox{$a\mp 1$}}&=&
\frac{\theta_1(2-u)\theta_1(3-u)}{\theta_1(2)\theta_1(3)},
\np
\raise 2mm \vtop{\hbox{$\quad a$}\hbox{$a\pm 1$}}
\> \framebox[0.4cm][c]{$u$} \>
\raise 2mm \vtop{\hbox{$a\mp 1$}\hbox{$\quad a$}}&=&
-\Bigl(  \frac{S_{a-1} S_{a+1}}{S_a^2} \Bigr )^{1/2}
\frac{\theta_1(u)\theta_1(1-u)}{\theta_1(2)\theta_1(3)},
\np
\raise 2mm \vtop{\hbox{$\quad a$}\hbox{$a\pm 1$}}
\> \framebox[0.4cm][c]{$u$} \>
\raise 2mm \vtop{\hbox{$a\pm 1$}\hbox{$\quad a$}}&=&
\frac{\theta_1(3-u)\theta_1(\pm 4a+2+u)}{\theta_1(3)\theta_1(\pm 4a+2)} \np
& &+
\frac{S_{a\pm 1}}{S_a}
\frac{\theta_1(u)\theta_1(\pm 4a-1+u)}{\theta_1(3)\theta_1(\pm 4a+2)}, 
\hbox{ for } \theta_1(\pm 4a+2) \ne 0, \np
 &=&\frac{\theta_1(3+u)\theta_1(\pm4 a-4+u)}{\theta_1(3)\theta_1(\pm4 a-4)} \np
 & &+
 \Bigl (
        \frac{S_{a\mp1}\theta_1(4)}{S_a \theta_1(2)}-
	    	  \frac{\theta_4(\pm 2 a-5)}{ \theta_4(\pm 2a+1)}
                      \Bigr )
		\frac{\theta_1(u) \theta_1(\pm 4a-1+u)}
		     {\theta_1(3) \theta_1(\pm 4a-4)}, \hbox{  otherwise }.
\end{eqnarray}
Here $\theta_{1,4}(x) =\vartheta_{1,4}(\lambda x, \tau)$, 
\begin{eqnarray*}
\vartheta_1(x,\tau) &=&2 q^{1/4}\sin x
\prod_{n=1}^{\infty}(1-2q^{2n} \cos 2x+q^{4n})(1-q^{2n}), \\
\vartheta_4(x,\tau) &=&
\prod_{n=1}^{\infty} (1-2q^{2n-1} \cos 2x+q^{4n-2})(1-q^{2n}),
\end{eqnarray*}
and $q=\exp(-\tau)$.
$\lambda$ is a parameter of the model specified below and 
$S_a$ denotes
$$
S_a=(-1)^a \frac{\theta_1(4a)}{\theta_4(2a)}.
$$
The model exhibits four different physical regimes depending on parameters,
\begin{itemize}
 \item regime 1. $0<u<3 , \,\lambda=\frac{\pi L}{4(L+1)},\, L\ge 2 $
 \item regime 2. $0<u<3, \, \lambda=\frac{\pi (L+2)}{4(L+1)},\,  L \ge 3$
 \item regime 3. $3-\frac{\pi}{\lambda}<u<0, \,
  \lambda=\frac{\pi (L+2)}{4(L+1)}, \, L \ge 3 $
 \item regime 4.  $3-\frac{\pi}{\lambda}<u<0,\, 
  \lambda=\frac{\pi L}{4(L+1)}, \, L \ge 2$
\end{itemize}
We are interested in  regimes 2 and 3.
The central charges and the scaling dimensions 
of the leading perturbation have been
evaluated in \cite{WNS1, WNS2, WPNS}.
Specializing to $L=3$, they read
\begin{itemize}
 \item regime 2. $c=\frac{1}{2}, \triangle=\frac{1}{16} $
 \item regime 3. $c=\frac{6}{5}, \triangle=\frac{15}{16}.$
\end{itemize}
Obviously, in connection with the Zamolodchiklov's
argument, regime 2 attracts more attention.
We will see, however, that both of them can be treated
on a same footing.

The one particle excitations are examined in \cite{MO97}.
Eight particles are identified and their masses 
are summarized by a single formula,
$$
m_j= \sum_a \sin(\frac{a \pi}{30}),
$$
where\\
\centerline{
\begin{tabular}{|l|l|}
\hline
$j$ &    set of allowed $a$'s \\
\hline 
1 (1)  &    $\{1,11 \}$ \\
2 (3)  &    $\{2,10,12 \}$ \\
3 (5)  &    $\{3,9,11,13\}$ \\
4 (7) &    $\{4,8,10,12,14\}$ \\
5 (8)  &    $\{5,7,9,11,13,15\}$ \\
6 (6) &    $\{6,8,12,14\}$  \\
7 (2)  &    $\{7,13 \} $  \\
8 (4) &    $\{6,10,14 \}$ \\
\hline
\end{tabular} }\\
\centerline{ Table 1. }

A number $k$ in the bracket means that it corresponds to the $k-$ th
light particle.
These exponents appear in many contexts,
diagonal scattering matrices, Weyl reflections with respect to
Coxeter element and so on \cite{Dorey}.
They will re-appear in a novel context later.
%
%
%
\section{ Quantum Transfer Matrix }

The row to row transfer matrix 
$T_{\hbox{\footnotesize{RTR}}}(u)$
is defined by 
$$
(T_{\hbox{\footnotesize{RTR}}}(u))^{\{b\}}_{\{a\}} =\prod_{j=1}^M 
\raise 2mm \vtop{\hbox{$b_j$}\hbox{$a_j$}}
\> \framebox[0.4cm][c]{$u$} \>
\raise 2mm \vtop{\hbox{$b_{j+1}$}\hbox{$a_{j+1}$}}.
$$
From the commutativity of $T_{\hbox{\footnotesize{RTR}}}(u)$ 
with different spectral parameters,
it is natural to define the Hamiltonian of 
associated 1D quantum chain by
$$
{\cal H}_{\epsilon} = \epsilon \frac{\partial}{\partial u} 
\ln T_{\hbox{\footnotesize{RTR}}}(u) |_{u=0}
$$
as in \cite{BWN}. 
Here $\epsilon=-1,(1)$ labels regimes 2 (3). 
In order to evaluate the free energy 
, we adopt the method of the quantum transfer matrix  
%
%
which is defined in the
following "staggered" manner,

$$
%
%
(T_{\hbox{\footnotesize{QTM}}}(u,v))^{\{b\}}_{\{a\}} =\prod_{j=1}^{N/2}
\mbox{\parbox[c][0.9cm]{0cm}{}}^{b_{2j-1}}_{a_{2j-1}}
\fbox{\parbox[c][0.7cm]{0.7cm}{ {\scriptsize $u\!+\! iv$}  }} \>
\mbox{\parbox[c][0.9cm]{0cm}{}}^{b_{2j}}_{a_{2j}}
\> \>
\mbox{\parbox[c][0.9cm]{0cm}{}}^{\,\,a_{2j+1}}_{a_{2j}}
\fbox{\parbox[c][0.7cm]{0.7cm}{ {\scriptsize $u\!-\! iv$}  }} \>
\mbox{\parbox[c][0.9cm]{0cm}{}}^{b_{2j+1}}_{b_{2j}}.
$$
Note that the number of sites $N$, sometimes
referred to as the Trotter number, has no relation to
the real system size $M$ and is even by the construction.
We further remark the commutative property of QTMs,
$$
[T_{\hbox{\footnotesize{QTM}}}(u,v), T_{\hbox{\footnotesize{QTM}}}(u,v') ]=0.
$$
The free energy per site is represented {\it only} by the
largest eigenvalue of $T_{\hbox{\footnotesize{QTM}}}$ at
$v=0$ and $u= -\epsilon \frac{\beta}{N}$,
\begin{eqnarray*}
\beta f &=&
-\lim_{M\rightarrow \infty} \frac{1}{M} 
   \ln {\hbox{Tr }} \exp(-\beta {\cal H}_{\epsilon})  \\
 &=& -\lim_{N\rightarrow \infty}\ln  \Bigl( \hbox{the largest eigenvalue of }
 T_{\hbox{\footnotesize{QTM}}}(u= -\epsilon \frac{\beta}{N}, v=0) \Bigr ).
\end{eqnarray*}
Explicitly, the eigenvalue $T_1(u,v)$ of $T_{\hbox{\footnotesize{QTM}}}(u,v)$
is given by
\begin{eqnarray}
T_1(u,v)&=& 
w \phi(v+\frac{3}{2}i) \phi(v+\frac{1}{2}i)\frac{Q(v-5/2 i)}{Q(v-1/2 i)} +
\phi(v+\frac{3}{2}i) \phi(v-\frac{3}{2}i) 
\frac{Q(v-3/2i)\, Q(v+3/2 i)}{Q(v-1/2 i)\, Q(v+1/2 i)} \np
&+ &w^{-1} \phi(v-\frac{3}{2}i) \phi(v-\frac{1}{2}i)  
 \frac{Q(v+5/2 i)}{Q(v+1/2 i)},  \\
Q(v)&:=&\prod_{j=1}^{N/2} h[v-v_j] \np
\phi(v) &:=& \Bigl(\frac{h[v+(3/2-u)i] h[v-(3/2-u)i]}{h[2i] h[3i]}\Bigr)^{N/2},
\qquad h[v] := \theta_1(iv),  \np
\end{eqnarray}
where  $w=\exp(i \pi \ell/(L+1)$ ($\ell=1$ for the largest eigenvalue
sector).
The parameters, $\{ v_j \}$ are solutions to "Bethe ansatz equation" (BAE),
\begin{equation}
w \frac{\phi(v_j+i)}{ \phi(v_j-i)}=
-\frac{Q(v_j-i) Q(v_j+2 i)}{Q(v_j+i)Q(v_j-2 i)},
\quad j=1,\cdots, \frac{N}{2}.
\label{bae}
\end{equation}
Now the difficulty arises. 
$N$ comes into
the expression of the free energy through $u$, as well as
the rhs of (\ref{bae}).
Thus  a simple-minded transformation of  BAE into
an integral equation using the root density function is not
 legitimate in contrast to $T=0$ problems.
A naive extrapolation to $N \rightarrow \infty$
by numerics may bring about errors.
One must thus devise other methods in encoding the
information of BAE roots.
In this report, we make use of the
commuting family of the transfer matrices (fusion hierarchy).
\footnote{
Results on other choices of auxiliary functions
 for several models, see \cite{Klu93,JKStj,JKS2p,JKSHub}}
Indeed, it has been known that 
the functional relations among them, which hold for
arbitrary $N$, can be an alternative
to BAE \cite{BP, KP}.
In the next section, we shall argue the
$sl_3$ fusion structure in  the dilute $A_L$ model.
%
%
\section{$sl_3$ fusion structure}
The $sl_3$ type fusion structure in the dilute $A_L$ model
has been discussed in details \cite{ZPG}.
This comes from the singularity of the RSOS weights at
$u=\pm 3$ where the face operator becomes projectors. 
A desired subspace can be picked up from tensor products
of spaces by using these projectors.
(For  explicit procedure, see\cite{ZPG}.)
The adjacency matrices are identified with Young diagrams, 
and combinatorics of tableaux  describe 
their tensor decomposition.

We are interested in eigenvalues of fusion QTMs.
In this view point, the most relevant fact is that 
these eigenvalues are again expressible in terms of 
"Young tableaux" depending on spectral parameters.
Let three boxes with
letters 1,2 and 3 represent the three terms in
eigenvalue of the quantum transfer matrix;
$$
T_1(u,v)= \framebox[0.4cm][c]{1}_{v} + \framebox[0.4cm][c]{2}_{v} +
\framebox[0.4cm][c]{3}_{v}. 
$$
Obviously, the eigenvalue of a fusion QTM can be represented by
sum over products of "boxes" with different letters and spectral parameters.
The point is that 
the assembly of such boxes can be identified with semi-standard 
Young tableaux (SST) for $sl_3$.
We present a simple example. (See \cite{BR,SuzG2,KS,KOS,AK,ZPG}
for details.)
By the fusion procedure, one can construct a transfer matrix 
of which auxiliary space acts on a symmetric subspace of 
$V \times V$. 
The set of the SST ,
$\framebox[0.4cm][c]{$i_1$}\framebox[0.4cm][c]{$i_2$}, (i_1\le i_2)$
is associated with this subspace.
The eigenvalue of the transfer matrix  is then represented by,
\begin{equation}
\sum_{i_1\le i_2} \framebox[0.4cm][c]{$i_1$}_{\,v-i} \,
\framebox[0.4cm][c]{$i_2$}_{\,v+i}.
\label{box-expr-rule}
\end{equation}
Similar construction leads to fusion models based on general Young tableaux,
of which eigenvalues are expressed by their shapes.
On each diagrams, the spectral parameter changes $+2i$ from left to right and
$-2i$ from top to the bottom. 
We restrict ourselves to rectangular shapes in this section.
There are still interesting properties, some of which
are crucial for the discussion in the next section.
First, due to identities,
\begin{eqnarray*}
\begin{tabular}{|l|}
\hline
1 \\
\hline
2 \\
\hline
\end{tabular}
\raise 2mm \vtop{  \hbox{$\phantom{}_{v+i}$} \hbox{$\phantom{}_{v-i}$}  }
&=& \phi(v+\frac{5}{2}i) \phi(v-\frac{5}{2}i) \framebox[0.4cm][c]{$1$}_{v},
\quad
\begin{tabular}{|l|}
\hline
1 \\
\hline
3 \\
\hline
\end{tabular}
\raise 2mm \vtop{  \hbox{$\phantom{}_{v+i}$} \hbox{$\phantom{}_{v-i}$}   }
=\phi(v+\frac{5}{2}i) \phi(v-\frac{5}{2}i) \framebox[0.4cm][c]{$2$}_{v}  \np
\begin{tabular}{|l|}
\hline
2 \\
\hline
3 \\
\hline
\end{tabular}
\raise 2mm \vtop{  \hbox{$\phantom{}_{v+i}$} \hbox{$\phantom{}_{v-i}$}   }
&=&\phi(v+\frac{5}{2}i) \phi(v-\frac{5}{2}i) \framebox[0.4cm][c]{$3$}_{v}
\qquad
\begin{tabular}{|l|l}
\hline
1\\
\hline
2 \\
\hline
3\\
\hline
\end{tabular}
\begin{array}{l}
 \phantom{}_{v+2i}  \\
 \phantom{}_{v}    \\
 \phantom{}_{v-2i}
\end{array}
=\prod_{j=1}^3 \phi(v+(\frac{9}{2}-j)i) \phi(v-(\frac{9}{2}-j)i),
\end{eqnarray*}
the QTMs from  $2\times m$ ( $3\times m$)  Young tableaux  reduce
to those from $1\times m$ (or just scalars).
Second, the eigenvalues of $1\times m$  fusion QTMs have the "duality"
in the sense of eq.(\ref{dualonerow}).
For the explanation, we introduce renormalized $1\times m$  fusion QTMs 
$T_m(v)$ by 
$$
T_m(v)= \frac{1}{f_m(v)} 
\sum 
\begin{tabular}{|l|l|l|l|}
\hline
$i_1$& $i_2$& $\cdots$& $i_3$ \\
\hline
\end{tabular}
$$
where the semi-standard (SS) condition $i_1 \le i_2 \le \cdots \le i_m$ is
imposed on the summation. 
(From now on we suppress the dependency on $u$ which must be
kept identical for all fusion QTMs.)
The spectral parameters, $v-i(m-1) \cdots v+i(m-1)$, 
are assigned  from left to right. 
A renormalization factor, which is the common factor of 
 the expressions from tableaux of length $m$, is given by
$$
f_m(v):= \prod_{j=1}^{m-1} \phi(v+i(\frac{2m-1}{2}-j)) 
\phi(v-i(\frac{2m-1}{2}-j)).
$$
Then the  resultant $T_m$'s are all degree $2N$ w.r.t. 
$h[v+\hbox{ shift }]$, and 
 have a periodicity due to Boltzmann weights;
$$
T_m(v+16/5 i) =T_m(v). 
$$
From the $sl_3$ structure, together with the above property,
the following functional relations are valid,
\begin{eqnarray}
T_m(v-i) T_m(v+i) &=&g_m(v) T_m(v) + 
          T_{m+1}(v) T_{m-1}(v),  \quad m \ge 1 \np
g_m(v) &=& \phi(v+i(m+3/2))  \phi(v-i(m+3/2)),	 \np
T_{-1}(v) &:=& 0   \np
T_0(v) &:=& f_2(v).
\label{sl3fusion}
\end{eqnarray}
The periodicity, $\phi(v + 16/5i)=\phi(v)$, leads to 
$g_{m+16}(v)=g_m(v), (g_{8+m}(v)=g_m(v + 8/5i)) \,\, m\ge 0$ 
and $g_{5-m}(v)=g_m(v \pm 8/5i), (0\le m\le 5) $.
As the adjacency matrices are vanishing, $T_6(v)=T_7(v)=0$.
Thus one deduces the duality relations,
\begin{equation}
T_m(v)=T_{5-m}(v + \frac{8 i}{5}), \,\, m=0,\cdots, 5
\label{dualonerow}
\end{equation}
and $T_{m+8}(v)=T_{m}(v+ \frac{8 i}{5}), m \ge -1$ 
for the solutions to eq.(\ref{sl3fusion}). 
These relations can be 
in principle proved by using explicit fusion weights.
We have at least checked the validity numerically.
Thus, for the dilute $A_3$ case, only  5 
"one-row" fusion QTMs need consideration.

Finally we discuss 
the closed set of functional relations among rectangular types,
motivated by success of several other cases.
From $T-$ system (\ref{sl3fusion}),
we are led to introduce the "${\cal Y}-$" functions 
as in \cite{KP,KNS2,Zhou}
$$
{\cal Y}_m(v) :=
\frac{T_{m+1}(v) T_{m-1}(v)}{g_m(v) T_m(v) }, (1\le m \le 4),
$$
which satisfy closed set of equation,
\begin{equation}
{\cal Y}_m(v-i) {\cal Y}_m(v+i)=
 \frac{(1+{\cal Y}_{m+1}(v)) (1+{\cal Y}_{m-1}(v))}
              {1+{\cal Y}_m(v)} (1\le m \le 4),
\label{recysys}
\end{equation}
where ${\cal Y}_5(v)=0$.

In \cite{Klu92, JKSfusion, KSS98}, TBA equations are re-derived from
such subsets of fusion hierarchy, as ${\cal Y}$ equations possess
good analytic properties;
both sides of equations are analytic and nonzero and have constant 
asymptotic behavior 
as $ |v| \rightarrow \infty $ (ANZC) in certain strips.
By taking the logarithmic derivative of both sides,
and applying Cauchy's theorem, one can directly rewrite 
 ${\cal Y}$ equations by  coupled integral equations which are nothing but
TBA equations.

The situation is different for the present model;
numerically, we find
"intrinsic" zeros and singularities of ${\cal Y}_m$'s 
in the strips $\Im v =[-1, 1]$ and $\Im v =[-1.6, -1]\cup [1,1.6]$.
Therefore the above argument can not be applied to (\ref{recysys})

Summarizing, the restriction to rectangle shapes
leads to a closed set of equations, but with 
poor analytic properties.

In the next section,
we thus consider a wider class of skew Young tableaux which
does lead to finitely closed functional relations having
the desired property.
%
%
%
\section{quantum Jacobi-Trudi formula and $E_8$ fusion hierarchy}
Let $\mu$ and $\lambda$ be a pair of Young tableaux satisfying 
 $\mu_i \ge \lambda_i, \forall i$.
We subtract  a diagram  $\lambda$ from $\mu$. 
The resultant "narrower" one, consisted of
$(\mu_1-\lambda_1,\mu_2-\lambda_2,\cdots )$ boxes is called 
a skew Young diagram $\mu-\lambda$.
(The usual Young diagram is the special case that $\lambda $ is empty, and
we will omit $\lambda$ in the case hereafter.)
\begin{figure}[hbtp]
\centering
  \includegraphics[width=8cm]{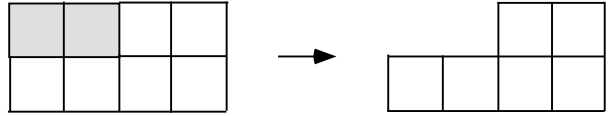}
\caption{An example of a skew Young Tableau, (4,4)-(2). In the lhs, 
the Young diagram (4,4) with
the shaded part (2) is given, while we have the corresponding 
skew diagram in the rhs.}
\label{skew1}
\end{figure}
In the theory of symmetric polynomials,  the Schur function
deserves the one of the fundamental objects. 
Generally,   
a Schur function is defined in correspondence to a 
skew Young diagram.
The Jacobi-Trudi formula asserts its representation 
 by a determinant of a matrix of which elements are
given by those  related to "one-row" diagrams
or " one-column" ones.
An analogous formula holds for the present situation
  \cite{BR, KS, KOS, AK}.

Consider a set of semi-standard skew Young 
tableaux of the shape $\mu-\lambda$.
As remarked in the previous section,
spectral parameters are assigned to boxes in a tableau
in such a manner that they change by $+2i$ from left to right and
$-2i$ from top to the bottom. 
We fix the spectral parameter of  the "top-left" box by  
$v+i(\mu_1-\mu_1')$
where $\mu'_1$ denotes the depth of the tableaux.
\begin{figure}[hbtp]
\centering
  \includegraphics[width=4cm]{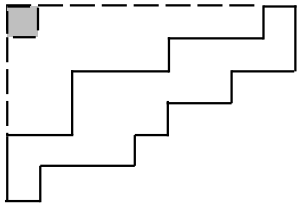}
\caption{The spectral parameter 
$v+i(\mu_1-\mu_1')$ is assigned to the hatched place }
\label{spectral}
\end{figure}
One regards each box in a tableau as an expression 
under the rule (\ref{box-expr-rule}) with
 the shift of the spectral parameter.
Then the product over all constituting boxes yields
an expression for a tableau.
\begin{thm}
Let ${\cal T}_{\mu/\lambda}(v)$ be the sum 
of the resultant expressions over the SST 
divided by a common factor,
$
\prod_{j=1}^{\mu'_1} f_{\mu_j-\lambda_j} 
  (v+i(\mu_1' -\mu_1+\mu_j+\lambda_j-2j+1)).
$
Then the following  equality holds.
\begin{equation}
{\cal T}_{\mu/\lambda}(v) =
  \hbox{det }_{1\le j,k\le \mu_1'} 
         ( T_{\mu_j-\lambda_k-j+k} 
            (v+i(\mu_1' -\mu_1+\mu_j+\lambda_k-j-k+1)) )
\label{qJT}
\end{equation}
where $T_{m<0}:=0$.
\end{thm}
We regard this as a quantum analogue of the Jacobi-Trudi
formula.
The proof utilizes the decomposition rules for
products of two tableaux with spectral parameters\cite{KS, KOS, AK}, which
is quite to parallel to those for Young tableaux.
One must only pay attention that the allowed positions of
a box is restricted by its spectral parameter.
${\cal T}_{\mu/\lambda}(v)$ may be naturally identified with 
the eigenvalue of QTM corresponding to fusion $\mu-\lambda$.
The proof of this is beyond the scope of the present report.
The important fact for our purpose is that 
 ${\cal T}_{\mu/\lambda}(v)$ defined in such a manner 
 is an analytic function of $v$
due to BAE, and contains  ${\cal T}_1(v)$ as a special case.
The former assertion is not obvious from 
the original definition by the tableaux, but
it is trivial from the quantum Jacobi-Trudi formula.

In the same spirit, we introduce  $\Lambda_{\mu/\lambda}(v)$, 
which is analytic under BAE,
  from ${\cal T}_{\mu/\lambda}(v)$ by putting
$T_{m\ge 6}(v)=0$ in the latter,
$$
 \Lambda_{\mu/\lambda}(v):=
 {\cal T}_{\mu/\lambda}(v) /.\{ T_{m\ge 6}(v) \rightarrow 0 \}.
$$
The pole-free property of $\Lambda_{\mu/\lambda}(v)$ is apparent from 
(\ref{qJT}).

The following "dualities" are simple corollaries of 
eq.(\ref{dualonerow}) and (\ref{qJT}).
\begin{lem}
\begin{eqnarray}
\Lambda_{(4,4)/(2)}(v) &=&
  \Lambda_{(3,1)}(v+\frac{2}{5}i),    \label{yd1} \\
\Lambda_{(4,2)/(1)}(v) &=&
  \Lambda_{(4,3)/(2)}(v+\frac{3}{5}i) ,  \label{yd2}\\
\Lambda_{(4,4)/(3)}(v) &=&
            \Lambda_{(4,1)}(v+\frac{7}{5}i),   \label{yd3}\\
\Lambda_{(7,4,4)/(3,3)}(v)&=&
          \Lambda_{(4,4,1)/(3)}(v+\frac{8}{5}i),   \label{yd4}\\
\Lambda_{(6,4,3)/(3,2)}(v)&=&
         \Lambda_{(5,4,2)/(3,1)}(v+\frac{8}{5}i),  \label{yd5} \\
\Lambda_{(7,7,4,4)/(6,3,3)}(v) &=&
    \Lambda_{(7,4,4,1)/(3,3)}(v+\frac{7}{5}i),   \label{yd6} \\  
\Lambda_{(7,7,4,4,1)/(6,3,3)}(v) &=&
  \Lambda_{(10,7,7,4,4)/(6,6,3,3)}(v+\frac{8}{5}i).   \label{yd7}\np        
\end{eqnarray}
\end{lem}
To establish a closed set of functional equations
we need further 
\begin{lem}.

The following relations hold,
\begin{eqnarray}
\Lambda_{(3,1,1)}(v) &=&\frac{f_5(v-2i)}{f_2(v-2i) f_2(v+i)} 
\Lambda_{(2)} (v+ 3i), 
\label{rel1}\\
\Lambda_{(6,4,4)/(3,3)}(v) &=&
\frac{f_5(v+i)}{f_2(v+i)} 
\frac{\Lambda_{(4,2)/(1)}(v-7 i)}{\phi(v-i5/2)\phi(v-i3/2) 
   \phi(v+i7/2) \phi(v+i9/2)} ,
\label{rel2} \\
\Lambda_{(10,7,7,4,4)/(6,6,3,3)}(v) 
&=& \phi(v-\frac{13}{2}i) \phi(v-\frac{7}{2}i)
     \phi(v+\frac{7}{2}i) \phi(v+\frac{13}{2}i) 
	   \Lambda_{(6,4,3)/(3,2)}(v),  
\label{rel3}   \\
 \Lambda_{(10,7,7,4,4,1)/(6,6,3,3)}(v) 
  &=& \phi(v-\frac{11}{2}i)\phi(v-\frac{5}{2}i)
       \phi(v+\frac{9}{2}i)\phi(v+\frac{15}{2}i)  \np
  & &  \Lambda_{(3,1)}(v-11i) \Lambda_{(4,3)/(2)}(v-2i) . 
\label{rel4}
\end{eqnarray}
\end{lem}
(\ref{rel1}) can be shown trivially from SS condition.
To prove the rest, it is convenient to go back to the
original definition of ${\cal T}$.
For example, we consider (\ref{rel2}).
\begin{eqnarray*}
{\cal T}_{(6,4,4)/(3,3)}(v)&=&\Lambda_{(6,4,4)/(3,3)}(v)+
        T_0(v)T_0(v+2i) T_8(v) \\
 &=&  \frac{f_5(v+i) f_3(v-5i) f_2(v+6i)}{f_2(v+i)f_3(v+5i) f_4(v-4i)} 
    T_3(v-5 i) T_2(v+6i)  \\
 &=& \frac{f_5(v+i)}{f_2(v+i)} \frac {T_3(v-5 i) T_2(v-10i)}
               {\phi(v-i5/2)\phi(v-i3/2) \phi(v+i7/2) \phi(v+i9/2)} .
\end{eqnarray*}
We have applied (\ref{qJT}) in the first line.
The second line follows from SS condition; 3 boxes in a 
column reduce to a scalar, 
so that the configurations in other two rows 
are completely independent.

\begin{figure}[hbtp]
\centering
  \includegraphics[width=8cm]{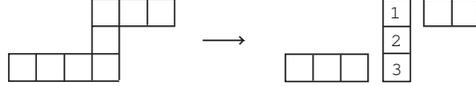}
\caption{Decomposition of the skew diagram (6,4,4)/(3,3).}
\label{decomp}
\end{figure}

In the last equation, the periodicity brings about
a new feature, namely, 
further decomposition of the products of $T$, 
$$
 T_3(v-5i) T_2(v-10i) =
 \Lambda_{(4,2)/(1)}(v-7 i)+ \frac{f_5(v-7i)}{f_2(v-10i) f_3(v-5i)} T_5(v-7i).
$$
Now that 
$T_8(v)=T_5(v)=T_0(v+\frac{8}{5}i)$ as discussed in the previous section, 
the proposition reduces to  
$$
T_0(v)T_0(v+2i) T_0(v+\frac{8}{5}i) 
= \frac{f_5(v-7i) f_5(v+i)}{f_2(v+i)f_3(v+5 i) f_4(v-4i)} T_0(v-7i+\frac{8}{5}i),
$$
which is proved by representing $T_0$ by $\phi$.
The other two relations are proved similarly.

We define the following fundamental objects (referred to as "modified fusion 
QTMs" in the introduction);
\begin{df}
\begin{eqnarray}
T^{(1)}(v) &:=& \Lambda_{(1)}(v) \np
T^{(2)}(v) &:=& \Lambda_{(4,1)}(v+\frac{7}{10}i)/\phi(v+\frac{8}{5}i) \np
T^{(3)}(v) &:=& \Lambda_{(7,4,4)/(3,3)}(v)/ 
                  \{ \phi(v+3/2i)\phi(v-3/2i)  \} \np
T^{(4)}(v) &:=& \Lambda_{(7,7,4,4)/(6,3,3)}(v+\frac{9}{10}i)/
             \{ \phi(v+i\frac{7}{5}) \phi(v+i\frac{8}{5})
			  \phi(v+i\frac{9}{5}) \} \np
T^{(5)}(v) &:=& \Lambda_{(5,4,2)/(3,1)}(v)  \np
T^{(6)}(v) &:=& \Lambda_{(4,2)/(1)}(v+\frac{13}{10}i) \np
T^{(7)}(v) &:=& \Lambda_{(2)} (v+\frac{8}{5} i)  \np
T^{(8)}(v) &:=& \Lambda_{(3,1)}(v-\frac{7}{10}i)/\phi(v+\frac{6}{5}i)
\label{defT}
\end{eqnarray}
\end{df}
In terms of $h[v+\hbox{ shift }]$, they are of degree,
$2N, 3N,4N,5N,6N,4N,2N,3N$, respectively.

\begin{prop}
The following "$E_8$ type T-system" holds among them,
\begin{eqnarray}
T^{(1)}(v-\dl i) T^{(1)}(v+\dl i) &=& \phi(v+i\frac{8}{5}) T^{(2)}(v)
               +T_0(v\pm \frac{9}{10}i) , \label{t1}\\
T^{(2)}(v-\dl i) T^{(2)}(v+\dl i) &=& T_0(v) T_0(v\pm \frac{8}{10}i)+
 T^{(1)}(v) T^{(3)}(v) ,    \label{t2}\\
T^{(3)}(v-\dl i) T^{(3)}(v+\dl i) &=&
T_0(v\pm \frac{1}{10} i)    T_0(v\pm \frac{7}{10}i)
+T^{(2)}(v) T^{(4)}(v),      \label{t3}\\
T^{(4)}(v-\dl i) T^{(4)}(v+\dl i) &=&
 T_0(v) T_0(v\pm \frac{1}{5}i) T_0(v\pm \frac{3}{5}i)+
 T^{(3)}(v) T^{(5)}(v),    \label{t4} \\
T^{(5)}(v-\dl i) T^{(5)}(v+\dl i) &=&
T_0(v \pm \frac{1}{10} i)T_0(v \pm \frac{3}{10}i)
T_0(v \pm \frac{5}{10}i) +
T^{(4)}(v) T^{(6)}(v)T^{(8)}(v) ,  \label{t5}\\
T^{(6)}(v-\dl i) T^{(6)}(v+\dl i) &=&
T_0(v \pm \frac{1}{5}i) T_0(v \pm \frac{2}{5}i)
+ T^{(5)}(v) T^{(7)}(v),  \label{t6}\\
T^{(7)}(x-\dl i) T^{(7)}(x+\dl i) &=&
T_0(x\pm\frac{3}{10}i) + T^{(6)}(x),  \label{t7} \\
T^{(8)}(v-\dl i)T^{(8)}(v+\dl i) &=&
T_0(v) T_0(v \pm \frac{2}{5}i) + T^{(5)}(v).  \label{t8} \np
\end{eqnarray}
\end{prop}
Here and in the following we use $g(x\pm y)$ to
 mean $g(x+y) g(x-y)$ for
some function $g$.

Note that the difference of the arguments of lhs is
much smaller ($\frac{1}{5}i$) than the difference
($2i$) in the original $sl_3$ type
fusion equation. 
This leads to the analytical property shown in the next
section.

Let us sketch the derivation of the above equations.
We consider the decomposition of $T^{(a)}$'s and their
dual partners.
The simplest case, (\ref{t1}), follows from consideration
on the decomposition of $T_1(v-4i) T_4(v+i)$.
From (\ref{qJT}) we have,
$$
T_1(v-4i) T_4(v+i) =\Lambda_{(4,1)}(v) +T_5(v) T_0(v-3i).
$$
Duality (\ref{dualonerow}) allows us to rewrite 
$T_4 (T_5)$ by of $T_1 (T_0)$.
After the shift $v\rightarrow v+0.7i$ together with periodicity,
(\ref{t1}) is established.
Eqs. (\ref{t2}), (\ref{t3}),(\ref{t4}) (\ref{t6}) and  (\ref{t7})
can be proved similarly.
Next we consider  (\ref{t5}) by utilizing,
\begin{eqnarray*}
& &\Lambda_{(7,7,4,4,1)/(6,3,3)}(v) 
        \Lambda_{(10,7,7,4,4)/(6,6,3,3)}(v+5i)  \\
& & = T_0(x\pm i) T_0(x\pm 9i) T_0(x+11 i)
         T_5(x\pm 4 i) T_5(x\pm 6i) T_5(x+14 i)  \\
& & +\Lambda_{(10,7,7,4,4,1)/(6,6,3,3)}(v+4i) 
        \Lambda_{(7,7,4,4)/(6,3,3)}(v+i).
\end{eqnarray*}
The eqs.(\ref{rel4}) and (\ref{yd2}) allow us to rewrite the rhs by 
$T^{(4)}, T^{(6)}$ and  $T^{(8)}$.
The second factor in the lhs can be rewritten
 by $T^{(5)}$ after transformations
$\Lambda_{(10,7,7,4,4)/(6,6,3,3)} \rightarrow  \Lambda_{(6,4,3)/(3,2)}$
((\ref{rel3}) in Lemma 2) $\rightarrow  \Lambda_{(5,4,2)/(3,1)}$ 
((\ref{yd5}) in Lemma 1).
The first term
can also be reduced to $T^{(5)}$ with one further step
((\ref{yd7}) in Lemma 1).
By shifting $v \rightarrow v-0.1 i$ and dividing both sides by
the factor 
$\phi(v\pm \frac{7}{5}i) \phi(v\pm \frac{8}{5}i) 
\phi(v\pm \frac{9}{5}i) \phi(v\pm 2i)$,
we obtain  (\ref{t5}).
(\ref{t8}) can be proved with the aid of (\ref{yd1}), 
(\ref{rel1}) and (\ref{rel2}).

\section{ $E_8$ Thermodynamic Bethe Ansatz from Y-system}
Now we transform the functional relations into "gauge
invariant" forms.
Let us define $Y-$ functions by combinations of $T^{(a)}$;
\begin{df}
\begin{eqnarray*}
Y^{(1)}(v) &:=& \frac{\phi(\renv+\frac{8}{5}i)T^{(2)}(\renv)}
                           {T_0(\renv\pm \frac{9}{10}i) }  \qquad
Y^{(2)}(v) :=\frac{T^{(1)}(\renv) T^{(3)}(\renv)}
                   {T_0(\renv) T_0(\renv\pm \frac{8}{10}i)}\\   
Y^{(3)}(v) &:=& \frac{T^{(2)}(\renv) T^{(4)}(\renv)}
                  {T_0(\renv \pm \frac{1}{10} i)T_0(\renv \pm \frac{7}{10}i)}   
                  \qquad
Y^{(4)}(v) :=\frac{T^{(3)}(\renv) T^{(5)}(\renv)}
               {T_0(\renv) T_0(\renv \pm \frac{1}{5}i)
               T_0(\renv \pm \frac{3}{5}i)}\\%
Y^{(5)}(v) &:=&\frac{T^{(4)}(\renv) T^{(6)}(\renv)T^{(8)}(\renv)}
  {T_0(\renv\pm \frac{1}{10} i)T_0(\renv\pm \frac{3}{10}i)
         T_0(\renv\pm \frac{5}{10}i)}
         \qquad
Y^{(6)}(v) := \frac{T^{(5)}(\renv) T^{(7)}(\renv)}
              {T_0(\renv\pm \frac{1}{5}i) T_0(\renv\pm \frac{2}{5}i)} \\
Y^{(7)}(v) &:=& \frac{T^{(6)}(\renv) }{T_0(\renv\pm\frac{3}{10}i) }
\qquad
Y^{(8)}(v) :=\frac{T^{(5)}(\renv)}{T_0(\renv) T_0(\renv \pm \frac{2}{5}i)}
\end{eqnarray*}
\end{df}
then  functional relations follow from the $T-$system.
To write it down neatly, we assign numbers to nodes in the
Dynkin diagram of $E_8$ as in the following figure.

\begin{figure}[hbtp]
\centering
  \includegraphics[width=8cm]{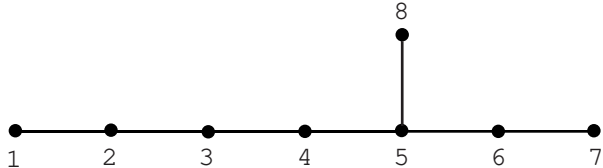}
\caption{The Dynkin diagram for $E_8$.}
\label{dynkin}
\end{figure}

We denote $a\sim b$ if $a$ and $b$ are adjacent in the Dynkin diagram.
\begin{thm}
Functional relations among $Y-$ functions exhibit the $E_8$ structure
in the following form, 
$$
Y^{(a)}(v- i)Y^{(a)}(v+ i) = \prod_{b\sim a} (1+ Y^{(b)}(v)), 
\qquad a=1,\cdots, 8.
$$
\end{thm}
This coincides with the $E_8$ case of the universal $Y-$ system
in \cite{AlZ}.
In order to reach our final goal, the TBA equation, 
further information is required on analytic structures of
$Y^{(a)}(v), 1+Y^{(a)}(v), a=1,\cdots,8$.
We employ numerical calculations for some 
fixed values of $q, N$ and $\beta$ for
this purpose.
Remark that only the largest eigenvalue
sector needs examinations.  
This is a drastic simplification from standard string hypothesis
argument in which analyses on all excitation sectors are, in principle, 
necessary.
Though we have performed numerics for relatively small values of $N$,
intriguing patterns are already found for zeros of $T^{(a)}(v)$.
First, these zeros are symmetric with respect to real axis,
which assures that $T^{(a)}(v)$'s are real on the axis.
Second, 
imaginary parts of coordinates
of zeros have the remarkable coincidence with 
the exponents in Table 1. 
See Table 2 for the 
example of the case $u=-0.08, q=0.3, N=12$. 
We summarize them as a conjecture for arbitrary $N$.
\begin{cj}
Zeros of $T^{(a)}$ distribute along  approximately on
the lines, $\Im v \sim \pm 0.1 (a_j+1)$. 
The set $\{a_j \}$ agrees with $\{a \}$ for the particle $j$ in Table 1.
\end{cj}
In addition, asymptotic behaviors are specified as,
\begin{cj}.
$Y^{(a)}$ 's have bounded asymptotic behaviors in the
largest eigenvalue sector of QTM.
Explicitly, their limiting values read,
\begin{eqnarray*}
Y^{(1)}(\asy) &=& 2(1+\sq),        \qquad Y^{(2)}(\asy)= (1+\sq)(5+3\sq), \\
Y^{(3)}(\asy)&=& 6(1+\sq)(3+2\sq), \qquad Y^{(4)}(\asy)= 4(4+3\sq)(5+3\sq), \\
Y^{(5)}(\asy)&=& 3(3+2\sq)^2 (5+4\sq), \qquad 
 Y^{(6)}(\asy)= 4(2+\sq)(4+3\sq), \\
Y^{(7)}(\asy)&=& 5+4\sq, \qquad Y^{(8)}(\asy)= 4(4+3\sq).
\end{eqnarray*}
\end{cj}
Combining these with the $T-$ system, we have
\begin{lem}.
Assume that  above conjectures are valid. 
Then $\widetilde{Y^{(a)}}(v)$ and  $1+Y^{(a)}(v)$ 
are Analytic, NonZero and have Constant asymptotic
behavior (ANZC) in  strips $\Im v \in [-1,1], [-0^+, 0^+]$,
respectively.
\end{lem}
$\widetilde{Y^{(a)}}(v)= Y^{(a)}(v)$ for $a \ne 1$ and,
$$
\widetilde{Y^{(1)}}(v)=
  Y^{(1)}(v) \{\kappa(v+i(1+\tilde{u})) 
               \kappa(v-i(1+\tilde{u}))\}^{\epsilon}
$$
where $\epsilon=1 (-1)$ for $u>0 (u<0)$ 
, $\tilde{u}=u/10$ and
$$
\kappa(v) =\Bigl( \frac{\vartheta_2(i\pi v/4, \tau')}
                   {\vartheta_1(i\pi v/4, \tau')} \Bigr )^{N/2}.
$$
Here $\tau'= 8 \tau$ is introduced so as to respect the periodicity
of $Y-$ functions along the real axis.
Remark that in deriving Lemma 3  we actually do not need such 
fine structures as given in Conjecture 1.  
Lemma 3 is robust in this sense.

Thanks to the simple identity $\kappa(v+i)\kappa(v-i) =\pm 1$,
the lhs of the theorem 2 can be replaced by
 $\widetilde{Y^{(a)}}(v-i) \widetilde{Y^{(a)}}(v+i)$.
Then we are in position to apply 
the Cauchy theorem directly and transform 
logarithmic derivatives of
resultant algebraic functional relations to a set of 
coupled integral equations.
Note that the resultant equations depend on
Trotter number only in the renormalization factor of $Y^{(1)}$
through the combination, roughly speaking, $u N$.
The limit $N\rightarrow \infty$ is thus performed straightforwardly.

Let us introduce  Fourier transformations  by 
$$
\widehat{F}(x) =\int^{16\tau/\pi}_{-16\tau/\pi}  e^{-ivx} F(v) dv, \qquad
F(v) = \frac{\delta}{2\pi} \sum_n e^{ivx_n} \widehat{F}(x_n)
$$
with $\delta=\pi^2/(16 \tau), x_n= n \delta$.
We also denote by $\widehat{\ln} F (x)$ the FT of logarithm of $F(v)$.

\begin{thm}
By Lemma 3, one can prove 
a set of nonlinear integral equations  among $\ln Y^{(a)}(v)$ 
for arbitrary $N$.

By taking the limit $N \rightarrow \infty$, the resultant equations 
read in the Fourier space as,
\begin{eqnarray}
\widehat{\ln} Y^{(a)} (x) &= &-\epsilon \delta_{a,1} 
  \widetilde{\beta } \widehat{s}(x)
     +  \widehat{ {\cal C}}_{a,b}(x) \widehat{\ln} (1+Y^{(b)}) (x)   \np
\widehat{s}(x) &=& \frac{1}{2 \cosh x}  \np
 \widehat{ {\cal C}}_{a,b}(x) &=& \widehat{s}(x) (2 I-{\cal C}^{E_8} )_{a,b},  
\end{eqnarray}
where $ \widetilde{\beta }=20\pi \beta$, 
and ${\cal C}^{E_8}$ denotes the Cartan matrix for $E_8$.
\end{thm}

The free energy is expressed via $Y-$ functions with the aid of (\ref{t1}).
\begin{eqnarray}
-\beta f &=& -\beta e_0 -\widetilde{\beta} b_1*s(0) +s*\ln (1+Y^{(1)})(0)
 \quad \epsilon=1, \\
     &=& \beta e_0-\widetilde{\beta} b_0*\rho_0 +
	        a_{1,t}^{E_8}*\ln(1+(Y^{(t)})^{-1})(0)
 \quad \epsilon=-1 \\
      e_0 &:=& \lambda [ \ln(\vartheta_1(\pi/16) \vartheta_1(3\pi/8) )]'  \np
     \widehat{b_0}(x)&: =&\frac{\sinh 6x}{\sinh 16x},
	     \quad 
	  \widehat{b_1}(x): =\frac{\sinh 5x + \sinh 15 x}{\sinh 16x},  \np
	  \widehat{\rho}_0(x) &:=& \frac{1}{2 \cosh x -1}, 
	     \quad
	  \widehat{a_{1,t}^{E_8}}(x):= \widehat{s}(x) 
	                \bigl( (1-\widehat{ {\cal C}}(x))^{-1} \bigr )_{1,t},
\end{eqnarray}
where $A*B:= \int^{16 \tau/\pi}_{-16 \tau/\pi} A(v-v') B(v') dv'$.
The final expressions are given in different forms depending on
the magnitudes of $Y$ in the vicinity of the
origin , ($|Y^{(a)}| \ll 1 (\gg 1)$ for $\epsilon =1 (-1)$).

The set of nonlinear integral equations is nothing but the
TBA equations obtained in \cite{BWN}, and the
free energy also coincides with their result.
(Note that 
the normalization of  our hamiltonian is $20\pi$ times the one in \cite{BWN}.)

Therefore, we have reached the same results, however, completely
independent of the string hypothesis.

\section{Summary and Discussion}

In this report we have applied the recent developed QTM method to
the 1D quantum chain related to the dilute $A_3$ model.
Modified QTMs described by skew Young tableaux have played a role, 
which has not been observed in former successful applications of
the QTM method.
The $E_8$ structure of the TBA equation 
has been  recovered without hypothesis on
specific forms of dominant solutions.
This gives further supports and consistency of the
"Trinity" among minimal unitary CFT theory $p=3$ perturbed by
$\phi_{1,2}$, the Ising model in a field and 
the  dilute $A_3$ model.

There may be a direct connection to $E_8$.
In the context of the analytic Bethe ansatz, the dressed vacuum
form for $E_8$ case , consisted of 249 terms,
is already obtained explicitly \cite{KSunp} as shortly remarked in \cite{KS}.
It would be interesting to show the direct reduction of
these 249 terms to {\it only} 3 terms.

The method employed here is in principle applicable to 
dilute $A_4$, $A_6$ model of which underlying 
symmetries are expected to be of 
$E_7$ and $E_6$.
Partial evidences are found very recently in a different
context \cite{BS3}.
We hope to report the study on them, as well as
 the extensive numerical studies on the $E_8$ case in
the near future.

\section*{Acknowledgments}
The author thanks Y.K. Zhou for discussions and
for making his unpublished results available.
He also thanks A. Kuniba for critical reading of the manuscript.

\newpage
\centerline{{\bf Table 2} } 
\vskip 0.9 cm
\begin{tabular}{|l|l|l|}
\hline
\phantom{}&   exponent&  locations of zeros \\
\hline 
$T^{(1)}$&  1  & $\pm 0.0160536139 \pm 0.194948867 i, 
                 \pm 0.0538522756 \pm 0.192860391 i,$ \\
         &     &  $\pm 0.124085873 \pm 0.179123531 i $ \\
\phantom{}&  11  &    $\pm 0.0161627786\pm 1.19901092i, 
                      \pm 0.0544016497 \pm 1.19859179i$,\\
          &  &             $ \pm 0.129454048 \pm 1.19555064 i$ \\
\hline
$T^{(2)}$& 2   &    $\pm 0.0160900147 \pm 0.295944203 i,
                     \pm 0.0540354627 \pm 0.294252469 i $\\
         &     &    $\pm 0.125806251 \pm 0.282759557 i  $ \\
\phantom{}&10  &    $\pm 0.0161506232 \pm 1.09758994 i, 
                     \pm 0.0543363518 \pm 1.09656338i$,\\
          &    &      $ \pm 0.128545339 \pm 1.08913466i$ \\
          &12 &       $\pm 0.0161758534 \pm 1.30002265 i, 
		                \pm 0.0544655721 \pm 1.30003507i$,\\
        & &            $ \pm 0.130002895 \pm 1.30016934i$ \\
\hline
$T^{(3)}$& 3   &    $\pm 0.0160996137 \pm 0.396199839 i, 
                   \pm 0.0540833902 \pm 0.394610881 i, $\\
         &     &    $\pm 0.126249978 \pm 0.383723324 i  $ \\
         & 9 & $\pm 0.0161379988 \pm 0.997143234 i, 
		           \pm 0.0542727134 \pm 0.995931521i$, \\
         & &           $\pm 0.12791469 \pm 0.987294006i$ \\
\phantom{}&11  &    $\pm 0.0161709223 \pm 1.19860006 i,
                     \pm 0.0544390469 \pm 1.19800087i$, \\
         & &           $\pm 0.129624985 \pm 1.19357709i$ \\
\phantom{}&13 &    $\pm 0.0161774898 \pm 1.40026185 i, 
                   \pm 0.0544730978 \pm 1.40037631i$, \\
         & &           $ \pm 0.130044745 \pm 1.40126236 i$ \\
\hline  
$T^{(4)}$& 4   &    $\pm 0.0161030779 \pm 0.496280551 i^*, 
                     \pm 0.054100446 \pm 0.49472356 i^*, $\\
         &     &    $\pm 0.126398274 \pm 0.484020597 i^*  $ \\
         & 8 &     $\pm 0.0161379988 \pm 0.896993334 i, 
		               \pm 0.0542433172 \pm 0.895721491i,$\\
          &  &        $\pm 0.127657169 \pm 0.886719531 i$ \\
\phantom{}&10  &    $\pm 0.0161605145 \pm 1.09815223 i, 
                       \pm 0.0543870319 \pm 1.09736514 i,$\\
        &   &         $\pm 0.129124614 \pm 1.09163793i$ \\
\phantom{}&12  &    $\pm 0.0161742694 \pm 1.29883901 i, 
                      \pm 0.0544557686 \pm 1.29834129i,$\\
        &    &        $\pm 0.12979684 \pm 1.29464801i$ \\
\phantom{}&14  &    $\pm 0.0161781274 \pm 1.50032712 i, 
                    \pm 0.0544760749 \pm 1.50046909i,$\\
        &    &        $ \pm 0.130062127 \pm 1.50155269 i$ \\
\hline
\end{tabular}

\begin{tabular}{|l|l|l|}
\hline
$T^{(5)}$& 5   &    $\pm 0.0161049087 \pm 0.596310956 i^*, 
                       \pm 0.054109231 \pm 0.594765257 i^*, $\\
         &     &    $\pm 0.126463632 \pm 0.584119659i^*$ \\
         & 7 &     $\pm 0.0161297776 \pm 0.796944088 i, 
		              \pm 0.0542322987 \pm 0.795652813i,$\\
        &    &        $\pm 0.127566253 \pm 0.786537192 i$ \\
\phantom{}&9  &    $\pm 0.0161552792 \pm 0.998001843 i, 
                    \pm 0.0543613631 \pm 0.99715346 i,$\\
        &    &        $\pm 0.128903306 \pm 1.09163793i$ \\
\phantom{}&11  &    $\pm 0.0161643894 \pm 1.19839097 i, 
                       \pm 0.054406539 \pm 1.19770481i,$\\
        &    &        $\pm 0.129330606 \pm 1.19269022i$ \\
\phantom{}&13  &    $\pm 0.0161753699 \pm 1.3989042 i, 
                     \pm 0.0544612157\pm 1.39843379i,$\\
       &      &       $ \pm 0.129847801 \pm 1.39493192 i$ \\
\phantom{}&15  &    $\pm 0.0161783577 \pm 1.59987872 i, 
                      \pm 0.0544773043\pm 1.59981987i,$\\
      &         &     $ \pm 0.130078989 \pm 1.59928832 i$ \\
\hline
$T^{(6)}$& 6   &    $\pm 0.0161138939 \pm 0.696491843 i, 
                      \pm 0.0541531069 \pm 0.695016384 i, $\\
         &     &    $\pm 0.126821874 \pm 0.684758808 i  $ \\
         & 8 &     $\pm 0.0161532542 \pm 0.897952415 i, 
		             \pm 0.0543515414 \pm 0.897084183i,$\\
        &  &          $\pm 0.128823242 \pm 0.890831951 i$ \\
\phantom{}&12  &    $\pm 0.0161656325 \pm 1.29845609 i,
                     \pm 0.0544127321 \pm 1.2977971 i,$\\
          &   &       $\pm 0.129390339 \pm 1.29296911 i$ \\
\phantom{}&14  &    $\pm 0.0161777507 \pm 1.49949385 i, 
                      \pm 0.0544740453 \pm 1.49927557i,$\\
         &      &     $\pm 0.130033087 \pm 1.49763549i$ \\
\hline
$T^{(7)}$& 7   &    $\pm 0.0161395827 \pm 0.797498832 i, 
                      \pm 0.0542838157 \pm 0.796443159 i, $\\
         &     &    $\pm 0.128192944 \pm 0.788942255 i  $ \\
         & 13 &     $\pm 0.0161693285 \pm 1.39904538 i, 
		             \pm 0.054432565 \pm 1.39863761i,$\\
        &  &          $\pm 0.129668227 \pm 1.3956389 i$ \\
\hline
$T^{(8)}$& 6  &    $\pm 0.0161211658 \pm 0.696762829 i,
                   \pm 0.0541902581 \pm 0.695400475 i ,$\\
         &     &    $\pm 0.127219855 \pm 0.685875782 i  $ \\
         & 10 &     $\pm 0.0161593273 \pm 1.09824048 i,
		              \pm 0.0543817684 \pm 1.0974928 i,$\\
         &      &     $\pm 0.129119115 \pm 1.09207157$ \\
         & 14 &     $\pm 0.0161762757 \pm 1.4990832 i, 
		             \pm 0.0544660786 \pm 1.49868902 i,$\\
         &      &     $\pm 0.129915909 \pm 1.49573858$ \\
\hline
\end{tabular} \\
\noindent The locations of zeros of fusion QTMs for
$N=12, q=0.3,u=-0.08$. The corresponding exponents of
Table 1 are given in the second column.\\
\noindent * For these cases, we have a problem in the convergence by the 
simple M{\"u}ller algorithm. These values must be thus taken to be rather approximate.

\end{document}